\documentclass[a4paper,11pt]{article}
\usepackage{pos}
\usepackage[noend]{algpseudocode}

\usepackage[per-mode = symbol]{siunitx}
\usepackage{graphicx}
\usepackage{algorithm}
\usepackage{algpseudocode}
\newtheorem{theorem}{Theorem}
\usepackage{adjustbox}
\usepackage{tikzducks} 
\usepackage{graphicx}

\usepackage{svg}
\usepackage{amsmath}
\usepackage{tabto}
\usepackage{amsthm}

\newcommand{\E} {\mathbb{E}}
\newcommand{\V} {\mathbb{V}}

\newcommand{\trace}{\mathsf{tr}}
\newcommand{\tr}{\mathsf{tr}}

\newcommand{\offdiag}{\mbox{offdiag}}
\newcommand{\mathC}{\mathbb{C}}

\newcommand{\lr}{\text{lr}}
\newcommand{\nnz}{\text{nnz}}

\title{MG-MLMC++ as a Variance Reduction Method for Estimating the Trace of a Matrix Inverse}
\ShortTitle{Variance Reduction for Estimating the Trace}

\author[a]{Andreas Frommer}
\author*[a]{Mostafa Nasr Khalil}

\affiliation[a]{Department of Mathematics, Bergische Universit\"at Wuppertal}

\emailAdd{khalil@uni-wuppertal.de}

\abstract{Hutchinson's method estimates the trace of a matrix function $f(D)$ stochastically using samples $\tau^Hf(D)\tau$, where the components of the random vectors $\tau$ obey an isotropic probability distribution. Estimating the trace of the inverse of a discretized Dirac operator or variants thereof have become a major challenge in lattice QCD simulations, as they represent the disconnected contribution to certain observables. The Hutchinson Monte Carlo sampling, however, suffers from the fact that its accuracy depends quadratically on the sample size, making higher precision estimation very expensive. Meyer, Musco, Musco and Woodruff recently proposed an enhancement of Hutchinson's method, termed \texttt{Hutch++}, in which the sample space is enriched by several vectors of the form $f(D)\zeta$, $\zeta$ a random vector as in Hutchinson's method. Theoretical analyses show that under certain circumstances the number of these added sample vectors can be chosen in a way to reduce the dependence of the variance of the resulting estimator from the number $N$ of samples from $\mathcal{O}(1/N)$ to $\mathcal{O}(1/N^2)$.

In this study we combine \texttt{Hutch++} with our recently suggested multigrid multilevel Monte Carlo approach. We present results for the Schwinger discretization of the $2$-di\-men\-si\-onal Dirac operator, revealing that the two approaches contribute additively to variance reduction. 
}

\FullConference{%
The 39th International Symposium on Lattice Field Theory,\\
8th-13th August, 2022,\\
Rheinische Friedrich-Wilhelms-Universität Bonn, Bonn, Germany
}

\begin{document}
\maketitle
	
\section{Introduction}

In this study, we consider the task of estimating the trace of the inverse of a large sparse matrix $D \in C^{n\times n}$, $\tr(D^{-1}) = \sum_i^{n} (D^{-1})_{ii}$. While this task arises in a variety of different fields, we focus 

on applications in Lattice QCD, where the disconnected fermion loop contribution to an observable is obtained from the trace of the inverse of the discretized Dirac operator, possibly after multiplication with certain $\gamma$-matrices; see \cite{SextonWeingarten1994}. The disconnected fermion loop contributions become increasingly important, as they cannot be neglected anymore given the accuracy of current state-of-the-art lattice simulations.   

Due to its sheer size, the $n \times n$ matrix $D^{-1}$ cannot be computed directly, and the only way to access information on the entries of $D^{-1}$ is through matrix-vector multiplications $D^{-1} \zeta$, i.e.\ via the solution of linear systems with matrix $D$.  This is where stochastic estimation techniques come into play, starting with Hutchinson's method \cite{hutchinson90}. Its key component is the use of random vectors  $\zeta \in \mathC^n$, whose components $\zeta_i$ obey an isotropic distribution, i.e.\ 
\begin{equation} \label{eq:random_props}
\E[|\zeta_i|^2] = 1, \enspace \E[\zeta_i\zeta_j] = 0 \mbox{ for } i,j=1,\ldots,n, i\neq j . 
\end{equation}
Typically, one takes the components to be identically independent distribution (i.i.d.) complex numbers $z$ with $\E[z] = 0$ and $\E[|z^2|] =1$. A prominent example is the Rademacher vectors, where $z$ is uniform in $\{-1,1\}$. Averaging $\zeta^H D^{-1} \zeta$ over $s$ independent random vectors $\zeta$ gives an unbiased estimator for the trace.  Algorithm~\ref{alg:plain_Hutch} shows how to proceed if a given relative target accuracy $\epsilon$ (actually: a confidence level of 68\% corresponding to the $1\sigma$ confidence interval if we rely on the law of large numbers) is to be achieved. 

\begin{algorithm}
\caption{plain Hutchinson}\label{alg:plain_Hutch}
\begin{algorithmic}[1]
\Require $D\in \mathC^{n\times n}$ nonsingular, $\epsilon$ relative accuracy 
\Ensure Approximation $\tau$ for $\tr(D^{-1})$
\For{$s=1,2,\ldots$ }
    \State generate next random vector $\zeta_{s}$ \Comment{$\zeta_s$ i.i.d.\ satisfying \eqref{eq:random_props}} 
    \State $\tau_s \gets \zeta_s^HD^{-1}\zeta_s$ \Comment{ solve linear system }
    \State $\tau = \frac{1}{s}\sum_{i=1}^s \tau_i$ \label{line:trace_hutch} \Comment{sample mean}
    \State $V = \frac{1}{s-1}\sum_{i=1}^s |\tau_i-\tau|^2$  \label{line:var_hutch} \Comment{sample variance}
     \If{$V/s \leq (\tau\epsilon)^2$} 
        \State \textbf{stop}
    \EndIf
\EndFor
\end{algorithmic}
\end{algorithm}

More precise theoretical results are known for special classes of matrices as exemplified by the following theorem from \cite{roosta2015improved}.

\begin{theorem}
Assume that the matrix $A$ is symmetric and positive semidefinite. Let $\tr^H(A)$ denote the Hutchinson estimator with $s$ samples which are Rademacher vectors. Let $\epsilon, \delta \in (0,1)$. Then, if
\begin{equation}
s \geq \frac{6}{\epsilon^2} \log\frac{2}{\delta}
\end{equation}
one has
\begin{equation}\label{eq:relerr}
    \mathbb{P}\left(\left|\tr_s^H(A) - \tr(A)\right| \leq \epsilon \tr(A)\right) \geq 1-\delta.
\end{equation}

\end{theorem}

The above theorem is a quantitative illustration of the crucial draw-back of Monte Carlo trace estimation: The accuracy increases only with the square root of the number of samples, which makes high accuracy samples practically infeasible unless modifications are found which reduce the variance substantially.   

In section~\ref{variance_reduct:sec} we will discuss the most common methods for variance reduction of the Hutchinson estimator based on projections. The recent Hutch++ algorithm presented in~\cite{hpp} fits into this category with a special choice for the projection subspace. We do not consider probing methods, which can be used additionally for variance reduction.

In section~\ref{mlmc:sec} we then first brief\/ly recall the multilevel Monte Carlo approach relying on a multigrid hierarchy for the matrix $D$, and then present a new approach which combines multigrid multilevel Monte Carlo with the Hutch++ idea. Numerical results for the Schwinger model will be reported in section~\ref{numerics:sec}.

\section{Variance Reduction via Projection} \label{variance_reduct:sec}

For Rademacher vectors, the variance of the 
Hutchinson estimator for $\tr(D^{-1})$ is given by $\tfrac{1}{2}\| \offdiag(D^{-1}+D^{-T}) \|_F^2 $, for $Z_4$-vectors it is $\| \offdiag(D^{-1}) \|_F^2$; see \cite{frommer2022multilevel}, e.g., and the heuristics underlying variance reduction techniques typically rely on just reducing $\| D^{-1} \|^2_F$.

\subsection{Deflation}

Deflation aims to ``remove'' a part from the operator which contributes most to the Frobenius norm. Using an oblique or orthogonal projector $\Pi$ on a yet to be determined $k$-dimensional subspace one splits $D^{-1} = (I-\Pi) D^{-1} + \Pi D^{-1}$ . Usually, $\tr(\Pi D^{-1})$ can be reduced to the trace of a $k \times k$ matrix which can be evaluated directly, and the Hutchinson estimator is used on $(I-\Pi) D^{-1}$. A summary of different choices for the deflating subspace can be found in \cite{Deflation_2017}.

Often, the deflating subspace is built from (approxmations to) small eigenmodes of $D$, i.e.\ large eigenmodes of $D^{-1}$.  Deflation will thus become increasingly inefficient if the number of large eigenvectors increases with the dimension of $D^{-1}$ (“volume dependence”). Actually, as is argued in \cite{Deflation_2017}, it can be advantageous to base deflation on singular triplets rather than eigenmodes. This is because the Frobenius norm is the $2$-norm of the vector of singular values, so deflating the $k$ largest singular triplets via a projection on the space spanned by the corresponding $k$ (right) singular vectors of $D$ sets the $k$ largest singular values of $D^{-1}$ to $0$ in $(I-\Pi)D^{-1}$.

\subsection{Exact Deflation}
With $(u_i,v_i,\sigma_i)$ denoting the singular triplets of $D$,
$Dv_i = \sigma_i u_i$, and the singular values $\sigma_i$ ordered increasingly, exact deflation uses the orthogonal projector 
$\Pi = V_k(U_k^H D V_k)^{-1}U_k^HD = V_kV_k^H $, where $U_k = [u_1|...|u_k], V_k = [v_1|...|v_k]$.  
Then the trace of $D^{-1}$ can be split as  
\begin{equation}\label{tr_def:eq}
    \tr(D^{-1}) = \tr((I-\Pi)D^{-1}) + \tr(\Pi D^{-1}).
\end{equation}
The first term in eq.~(\ref{tr_def:eq}) can be expected to have reduced variance and can be estimated stochastically via Alg.~\ref{alg:plain_Hutch} with less samples. The second term is available directly since $\tr(\Pi D^{-1}) = \tr(V_k^HD^{-1}V_k) = \sum_{i=1}^{k} \frac{1}{\sigma_i}u_i^Hv_i$. If instead $U_k$ and $V_k$ contain the left and right eigenvectors belonging to the smallest eigenvalues $\lambda_i$ of $D$, then the oblique projector $\Pi =  V_k(U_k^H D V_k)^{-1}U_k^H D = V_kU_k^H$ achieves $\tr(\Pi D^{-1}) = \tr(U_k^H D^{-1} V_k) = \sum_{i=1}^k \frac{1}{\lambda_i}.$  
If $D$ is Hermitian and positive definite, the two deflation approaches coincide, since then left and right eigenvectors as well as left and right singular vectors all coincide, and the singular values are the eigenvalues.

\subsection{Inexact Deflation}
Exact deflation requires the precise computation of singular triplets or eigenpairs, which can be quite costly. We can instead work with approximations and still build the projection $\Pi$ the same way as in exact deflation. Now, $\tr(\Pi D^{-1})$ is not directly available from approximate singular triplets or eigenvalues and the projector  $V_k(U_k^H D V_k)^{-1}U_k^HD$ differs from
the projector $V_kV_k^H$, e.g. Using the former gives $\tr(\Pi D^{-1}) = \tr(V_k(U_k^H D V_k)^{-1}U_k^H)$, which requires the inversion of a small $k \times k$ matrix and $k$ multiplications with $D$. Using the latter gives 
$\tr(\Pi D^{-1}) = \tr(V_k^H D^{-1}V_k)$ which requires $k$ system solves with the large matrix $D$. 

If we have a sparse representation for $U_k$ and $V_k$, we can efficiently use very large values for $k$ in inexact deflation. This is the case with multigrid prolongation and restriction operators; see \cite{Balietal2015,frommer2022multilevel,romero2020multigrid} and section~\ref{mlmc:sec}.

\subsection{Hutch++}

Hutch++ \cite{hpp} is an inexact deflation method, where the deflating subspace is obtained from $D^{-1}$-images of random vectors: 

We precompute $y_i :=D^{-1}s_i$ for $d$ i.i.d.\ isotropic random vectors $s_i$. This is one step of a block power method to approximate the largest eigenpairs of $D^{-1}$. We 
build an orthogonal projector $\Pi$  on the space spanned by the $y_i$ as $\Pi = QQ^H$ with the columns of $Q \in \mathC^{n \times d}$ representing an orthonormal basis for that space spanned, typically obtained through a QR-factorization of $ Y= [y_1| \cdots |y_d]$. The range of the vectors $y_i$ contains, with increasing probability as $d$ increases, good approximations to eigenvectors belonging to large eigenvalues of $D^{-1}$. 
As before, we decompose the matrix as 

\begin{equation}\label{eq:hutchpp_decomp}
D^{-1} =  (I-Q) D^{-1} + Q D^{-1}. 
\end{equation}

As usual, the trace of the first summand in eq.\ (\ref{eq:hutchpp_decomp}) is estimated stochastically and should have a reduced variance. For the trace of the second term, we use $\tr(QD^{-1}) = \tr(V^H D^{-1} V)$, which requires another $d$ system solves with $D$. Under the assumption that a system solve with $D$ has cost $\mathcal{O}(n^2)$, an asymptotic analysis in \cite{hpp} shows that for a given budget of $N$ system solves, the optimal choice for $d$ is $d = N/3$. The recent paper \cite{adaptive_hutch++} develops an adaptive technique 
to choose $d$ optimally for a given target accuracy. All these results rely on the matrix being Hermitian (and positive definite). 

\section{Multilevel Monte Carlo (MLMC)}
\label{mlmc:sec}
MLMC \cite{Giles,Giles2015} is a generalization of standard Monte Carlo. The idea is to represent a random variable $X$ as a sum 
\begin{equation} \label{mlmc:eq}
X = \sum_{\ell=1}^L X_\ell 
\end{equation}
using additional random variables $X_\ell$ such that the variance of the $X_\ell$ is small when it is costly to evaluate and possibly large when it is cheap to evaluate. The different random variables can now be estimated stochastically and independently to obtain an estimator for $X$. 

The variance $\rho^2$ for the resulting estimator for $\E[X]$ is the sum of the variances of the estimators for $\E[X_\ell]$. In the {\em uniform} approach one chooses the number $N_\ell$ of samples at each `level' $\ell$ such that $\V[X_\ell]/N_\ell = \rho^2/L$. If one knows the cost $C_\ell$ for  an evaluation of $X_\ell$, the problem of minimizing the total cost under the constraint to obtain a variance of $\rho^2$ 
is solved for the {\em optimal} values \cite{Giles2015}
\begin{equation}\label{eq:accuracy_opt}
N_\ell = \frac{1}{\rho^2} \sqrt{\V[X_\ell] / C_\ell} \sum_{j=1}^{L-1} \sqrt{\V[X_j] C_j}. 
\end{equation}
The variance of the estimator for $X_\ell$ with $N_\ell$ samples is then
\begin{equation} \label{optimal_variance:eq}
\V[X_\ell] / N_\ell = \rho^2 \sqrt{\V[X_\ell] C_\ell} \; \left/ \; \sum_{j=1}^{L-1} \sqrt{\V[X_j] C_j}\right. .
\end{equation}

\subsection{Multigrid Multilevel Monte Carlo (MG-MLMC) for the trace}

In \cite{frommer2022multilevel} we proposed a multilevel Monte Carlo method based on a multigrid hierarchy to reduce the variance. 
One splits the original matrix $D_1^{-1} = D^{-1}$ into a telescopic sum as:
\begin{eqnarray} \label{mlmc_decomp_matrix:eq}
   D_1^{-1}\!
   &=& \! (D_1^{-1} - P_1 D_2^{-1} R_1) + (P_1 D_2^{-1} R_1 - P_1 P_2 D_3^{-1} R_2 R_1)  \ldots \nonumber 
   \mbox{}+ P_1\cdots P_{L-1} D_L^{-1} R_{L-1}\cdots R_1 \nonumber \\ 
        \! &=& \! \sum_{\ell = 1}^{L-1} \left( \hat{P}_\ell D_\ell^{-1} \hat{R}_\ell - \hat{P}_{\ell+1} D_{\ell+1}^{-1} \hat{R}_{\ell+1}  \right) + \hat{P}_{L} D_{L}^{-1} \hat{R}_{L}, \label{matrix_dec:eq}
\end{eqnarray}
where $
\hat P_\ell = P_1 \cdots P_{\ell-1}, \enspace \hat R_\ell = R_{\ell-1} \cdots \hat R_1. $

Here, the $P_\ell$ and $R_\ell$ are the prolongation and restriction operators
between consecutive levels of the multigrid hierarchy, respectively, $D_{\ell+1} = R_\ell D_\ell P_\ell$ are the (Galerkin) coarse grid operators, and $\hat P_\ell$ and $\hat R_\ell $ are the accumulated prolongations and restrictions which transport between level 1 and $\ell$. Note that with the projector $\Pi_1 = P_1D_2^{-1}R_1D_1$ we have $\Pi_1 D_1^{-1} = P_1D_2^{-1}R_1$ and similarly for the coarser levels, thus establishing the connection with inexact deflation discussed in section~\ref{variance_reduct:sec}. In multigrid, the prolongations $P_\ell$ are precisely constructed in a manner that they contain good approximations to the small eigenmodes or singular triplets of $D_\ell$.

The decomposition eq.~(\ref{matrix_dec:eq}) gives a multilevel decomposition for the trace as 
\begin{equation} \label{mlmc_decomp_trace:eq}
   \tr\left(D_1^{-1}\right)
   =  \sum_{\ell = 1}^{L-1} \tr \left( \hat{P}_\ell D_\ell^{-1} \hat{R}_\ell - \hat{P}_{\ell+1} D_{\ell+1}^{-1} \hat{R}_{\ell+1}  \right) + \tr\left(\hat{P}_{L} D_{L}^{-1} \hat{R}_{L}\right)
\end{equation}
to be used in a MLMC method. We expect the variance for each level difference $\hat{P}_\ell D_\ell^{-1} \hat{R}_\ell - \hat{P}_{\ell+1} D_{\ell+1}^{-1} \hat{R}_{\ell+1}$ to be small, since the prolongations $P_{\ell+1}$ are built to approximate small eigenpairs or singular triples of $D_\ell$. The sizes of the matrices to invert on each level difference decrease significantly with the level, thus making a stochastic sample increasingly less costly.

On the coarsest level $L$, depending on the size of the matrix $D_L$, we might  be able to compute the trace directly as $\sum_{i=1}^{N_L} e_i^T D_L^{-1}\hat R_L \hat P_L e_i$. If we do it stochastically, we have to invert a matrix whose dimension is very small compared to that of $D$. 

In the successful multigrid approaches for the Wilson-Dirac matrix or its twisted mass variant, see \cite{MGClark2007,MGClark2010_1,FroKaKrLeRo13,Bacchioetal2016}, the restrictions and prolongations are aggregation based with $R_\ell = P_\ell^H$, and their columns are orthonormal, $P_\ell^H P_\ell = I$. This is why, using the cyclic property of the trace, eq.~(\ref{mlmc_decomp_matrix:eq}) gives 
\begin{equation} \label{mlmc_dec_trace_reduced:eq}
        \tr(D_1^{-1}) = \sum_{\ell = 1}^{L-1}\tr\left(D_{\ell}^{-1} - P_{\ell} D_{\ell+1}^{-1} P_{\ell}^H\right) + \tr \left( P_{L-1} D_{L}^{-1} P_{L-1}^H \right).
\end{equation}

In contrast to eq.~(\ref{mlmc_decomp_trace:eq}) this allows to work with random vectors of  the smaller size $n_\ell$ instead of $n$ on the various difference levels.  

\subsection{Multigrid Multilevel Monte Carlo++ (MG-MLMC++)}
The idea of MG-MLMC++ is to apply the Hutch++ estimator for each of the level differences in the multilevel decomposition eq.~(\ref{mlmc_decomp_trace:eq}). We describe the method in Algorithm~\ref{alg:mlmcpp_optimal}.

\begin{algorithm}
\caption{MG-MLMC++, optimal accuracies}\label{alg:mlmcpp_optimal}
\begin{algorithmic}[1]
\Require $D\in \mathC^{n\times n}$ nonsingular, $\epsilon$ relative accuracy, $L$ number of levels, $\hat{R}_\ell, \hat{P}_\ell$ restriction and prolongation operators between levels 1 and $\ell$, $D_\ell \in \mathC^{n_\ell \times n_\ell}$ matrix on level $\ell$, $d_\ell$ number of deflation vectors on level $\ell$, $\ell = 1,\ldots, L$,  
\Ensure Approximation $\tau + \tau_L$ for $\trace(D^{-1})$
\For{$\ell = 1,\ldots, L-1$}  \Comment{obtain deflation vectors}
       \State generate $d_\ell$ i.i.d.\ random vectors $s_i, i=1,\ldots,d_\ell$,  \Comment{with distribution satisfying \eqref{eq:random_props}}
       \State collect them as columns in $S_\ell \in \mathC^{n\times d_\ell}$
       \State $Y_\ell \gets \left(\hat P_\ell D^{-1}_\ell \hat  R_\ell -  \hat{P}_{\ell+1} D^{-1}_{\ell+1} \hat R_{\ell+1}\right) S_\ell, $ \Comment{$Y_\ell \in \mathC^{n\times d_\ell}$, , $2d_\ell$ mg solves.}
       \State Compute QR-factoriz.\ $Y_\ell = Q_\ell K_\ell$ \Comment{$Q_\ell =[q_1|\cdots| q_{d_\ell}] \in \mathC^{n \times d_\ell}$ has orthon.\ cols}     
       \State $\tau_{\ell}^\lr \gets \sum_{i=1}^{d_\ell} q_i^H\left(\hat P_\ell D^{-1}_\ell \hat  R_\ell -  \hat{P}_{\ell+1} D^{-1}_{\ell+1} \hat R_{\ell+1}\right)q_i$ \Comment{\textit{low rank part}, use mg to solve lin.\ sys.}
\EndFor
\State $\tau_L \gets \sum_{i=1}^{N_L} (e_i^H\hat{P}_L)D_L^{-1}(\hat{R}_Le_i)$  \Comment{coarsest level is computed directly }
\State Set all levels $\ell$ to active   \Comment{non active levels have reached required accuracy} \label{line:set_lev}
\For{$s=1,2,\ldots$ \textbf{until} all levels $\ell$ not active} \Comment{\textit{stochastic part}}
  \For{$\ell = 1,\ldots, L-1$ \textbf{and} $\ell$ is active} 
        \State generate next random vector $\zeta_s$ \Comment{$\zeta_s$ i.i.d.\ satisfying \eqref{eq:random_props}}
        \State $z_s = \zeta_s-Q_\ell (Q_\ell^H \zeta_s)$ \label{line:zs} \Comment{projected vector}
        \State $\tau_{s,\ell} \gets z_s^H \left(\hat P_\ell D^{-1}_\ell \hat  R_\ell \zeta_s -  \hat{P}_{\ell+1} D^{-1}_{\ell+1} \hat R_{\ell+1}\zeta_s\right)$ \label{line:tau_s_ell}
        \State $C_{s,\ell} \gets$ cost for lines~\ref{line:zs} - \ref{line:tau_s_ell} 
        \State $\tau_\ell = \frac{1}{s}\sum_{i=1}^s \tau_{i,\ell}$, $V_\ell = \frac{1}{s-1}\sum_{i=1}^s |\tau_{i,\ell}-\tau_{\ell}|^2$   \Comment{sample mean and variance} 
 
         \State $C_\ell = \frac{1}{s} \sum_{i=1}^s C_{i,\ell}$   \Comment{average cost per sample}
    \EndFor
    \State $\tau = \sum_{\ell=1}^{L} (\tau_\ell + \tau_{\ell}^\lr)$
    \For{$\ell = 1,\ldots, L-1$ \textbf{and} $\ell$ is active}  \Comment{update target accuracies $\rho_\ell$}
        \State $\rho_\ell \gets \left(\sqrt{C_\ell V_\ell}\, / \, \sum_{j =1}^{L-1}\sqrt{C_jV_j} \right)^{1/2} \cdot (\epsilon \tau)$   
        \If{$V_{\ell}/s \leq \rho_\ell^2$} 
            \State set level $\ell$ to inactive
        \EndIf
    \EndFor
\EndFor \label{line:end_for}
\end{algorithmic}
\end{algorithm}

\pagebreak 
Some of its more important features are:
\begin{itemize}
\item We assume that we have a cost model to measure the cost for a stochastic sample. We take averages of the cost for each stochastic sample to get increasingly accurate average costs $C_\ell$.
\item With this measured cost and the measured sample variance $V_\ell$ we determine the optimal target variance from eq.~(\ref{optimal_variance:eq}) for each level difference. This target variance is updated at each additional sample on that level difference.
\item We describe the algorithm using the decomposition eq.~(\ref{mlmc_decomp_trace:eq}) with the accumulated prolongations and restrictions. The adaptation to eq.~(\ref{mlmc_dec_trace_reduced:eq}), should it apply, is straightforward.
\item The number of deflation vectors $d_\ell$ for each level difference must be chosen a priori.
\item Lines 5 and 6 perform one step of the block power iteration, the crucial ingredient of the Hutch++ method. We can perform more than 1, $k$ say, iterations of the block power method by repeating these lines with $S_\ell$  in the next sweep equal to $Q_\ell$ from the previous sweep. 
\end{itemize}

\section{Numerical Results}
\label{numerics:sec}
Numerical computations were performed using Python on a single core of a node with 44 cores Intel(R) Xeon(R) CPU E5-2699 v4 @ 2.20GHz. 
We demonstrate the benefits of MG-MLMC++ over exactly deflated Hutchinson and the benefits of MG-MLMC with the two types of accuracies by using the Schwinger discretization of the $2$-di\-men\-si\-onal Dirac operator \cite{Schwinger1962}
with the same configuration and parameters as in~\cite{frommer2022multilevel}. In particular, we use 5 different (negative) masses $m$ to shift the mass-less Schwinger operator by the respective multiple of the identity, thus yielding operators with increasing condition number. The multigrid hierarchy was constructed with a bootstrap setup and aggregation based prolongations as in DD$\alpha$AMG \cite{FroKaKrLeRo13}. Properties of the matrices at the various levels are summarized in the top part of Table~\ref{schwinger:tab}.

\begin{table}     
    \centering
    \begin{small}
    \begin{tabular}{|r|l|cccc|}
    \hline
    \multicolumn{6}{|c|}{Schwinger model} \\
      \hline
     $L$  &                &$\ell = 1$     &$\ell = 2$    &$\ell = 3$    &$\ell = 4$    \\ \hline 
     4   & $n_\ell$      & $2\cdot 128^2$& $4\cdot 32^2$& $8\cdot 8^2$ & $8\cdot2^2$  \\
          & $\nnz(D_\ell)$ & $2.94e5$       & $1.64e5$     & $2.46e4$     & $1024$       \\ \hline \hline
       mass   & \multicolumn{1}{c}{$m_1 = -0.1320$}      & $m_2= -0.1325$     & $m_3 = -0.1329$    & $m_4 = -0.1332$    &  $m_5 = -0.1333$   \\
defl. vects. & \multicolumn{1}{c}{$384$}  & $384$         & $512$        & $512$        & 512            \\ \hline     
    \end{tabular}
    \end{small}
        \caption{Parameters used in the Schwinger model and number of deflated eigenvectors chosen in exactly deflated Hutchinson. nnz($D_\ell$) denotes the number of non-zero elements in $D_\ell$}
 \label{schwinger:tab}
 \end{table}
 
To assess the performance of the algorithms we use a simple cost model which counts the arithmetic operations in all occurring matrix-vector multiplications, i.e.\ in the projections, the restrictions and prolongations and in the smoothing iteration in the multigrid solver. This arithmetic cost is proportional to the number of non-zeros in the respective matrix, and as an indication, this number is reported for the operators at the different levels in Table~\ref{schwinger:tab}.

We use a deflated Hutchinson method as our reference for comparison. We did not use non-deflated Hutchinson, because its performance is by two orders of magnitude worse than that of deflated Hutchinson. For deflation, we used the $k$ smallest eigenmodes that we precomputed, and then optimized $k$ so as to obtain the smallest overall cost, {\em excluding} the cost for the eigenvector computation. So the work for deflated Hutchinson is actually higher than what we report. 

\begin{figure}
\centering
\includegraphics[width=0.75\textwidth]{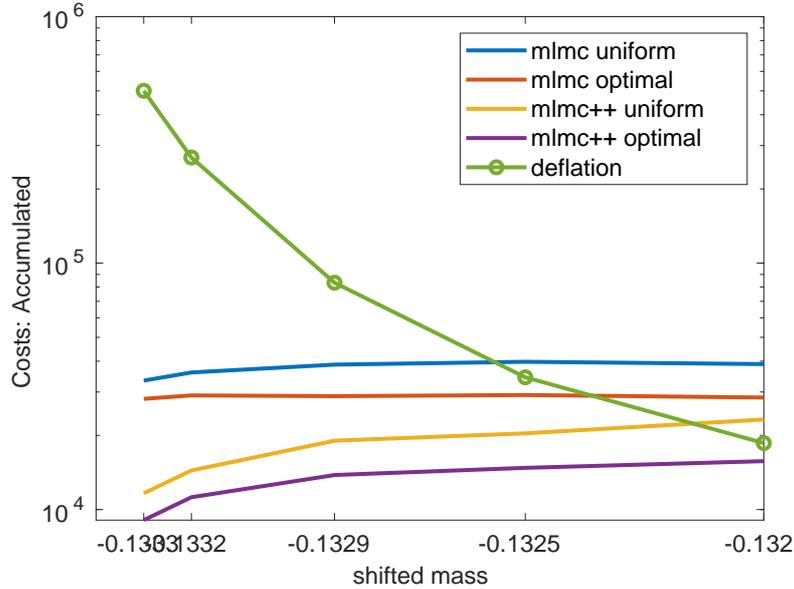}

 \caption{MG-MLMCM, MG-MLMC++ and deflated Hutchinson for the Schwinger matrix: total cost for different masses with uniform and the optimized target variances. 
 \label{fig:total_work}}
\end{figure}

\begin{figure}
\centering
\includegraphics[width=0.75\textwidth]{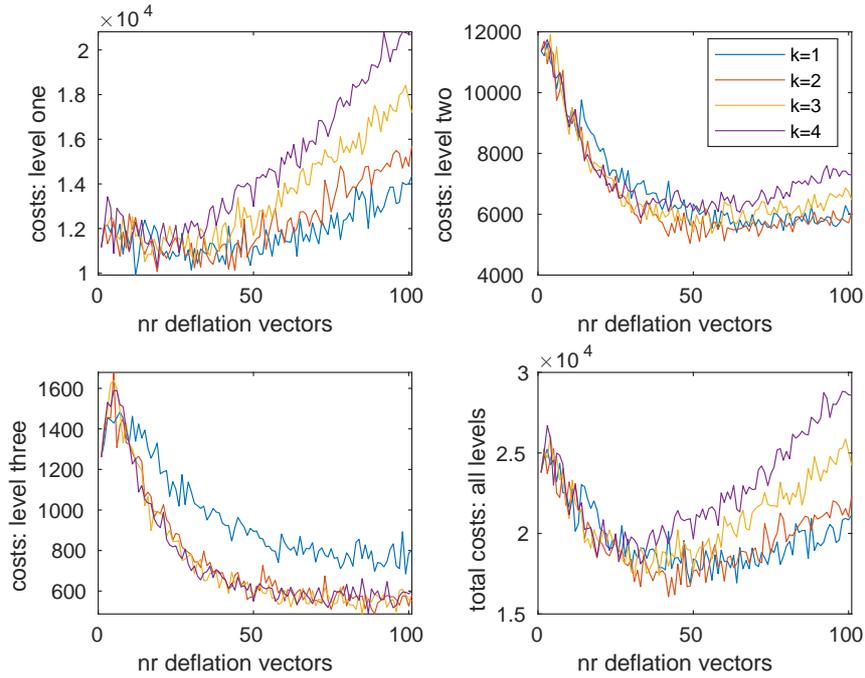}
 \caption{Work at each level difference as a function of the number of vectors in the block power iteration and total work when taking the same number on all levels. 
 . \label{fig:deflated_work}}
\end{figure}

Fig.~\ref{fig:total_work} reports the arithmetic cost in MFlops for five different methods: Deflated Hutchinson for reference, MG-MLMC with uniform target variances on the difference levels and its modification working with optimal target variances, and then the corresponding two versions for MG-MLMC++. Here, we determined the number $k$ of steps of the block power iteration and the number $d_\ell$ of vectors to be used there by a parameter scan on each level. This scan is reported in Fig.~\ref{fig:deflated_work}. We find that $k=2$ is 
 a better choice than $k=1$, and that increasing $k$ further does not result in significant further gains. Also, $d_\ell \approx 50$ appears as a good choice on all level differences.    

The plot in Fig.~\ref{fig:total_work} shows that for all masses considered, the best MLMC method now outperforms deflated Hutchinson (with an optimal number of deflated vectors and without counting the work for computing those). It also shows that with optimal numbers of vectors in the block power iteration, the ``++''-enhancement improves MLMC by a factor of $1.5$ to $3$, with a stronger improvement for the smaller values of $m$, i.e.\ the more ill-conditioned matrices. The influence of the strategy to determine the target variance (``uniform'' or ``optimized'') is, on the other hand, not very significant.

\begin{table}
\centering
\begin{tabular}{|l|lllll|l}
\cline{1-6}
method type & \multicolumn{5}{c|}{samples nr. per mass} &  \\ \cline{2-7} 
 & \multicolumn{1}{l|}{m1} & \multicolumn{1}{c|}{m2} & \multicolumn{1}{l|}{m3} & \multicolumn{1}{l|}{m4} & \multicolumn{1}{c|}{m5} & \multicolumn{1}{l|}{level} \\ \hline \hline
deflated Hutchinson & \multicolumn{1}{l|}{529} & \multicolumn{1}{l|}{1004} & \multicolumn{1}{l|}{2318} & \multicolumn{1}{l|}{7431} & 13845 & \multicolumn{1}{l|}{} \\ \hline \hline
MG-MLMC, optimized target variances & \multicolumn{1}{l|}{325} & \multicolumn{1}{c|}{321} & \multicolumn{1}{l|}{315} & \multicolumn{1}{l|}{313} & 306 & \multicolumn{1}{l|}{$\ell=1$} \\ \cline{2-7} 
 & \multicolumn{1}{l|}{854} & \multicolumn{1}{l|}{873} & \multicolumn{1}{l|}{837} & \multicolumn{1}{l|}{833} & 791 & \multicolumn{1}{l|}{$\ell=2$} \\ \cline{2-7} 
 & \multicolumn{1}{l|}{4208} & \multicolumn{1}{l|}{4218} & \multicolumn{1}{l|}{4414} & \multicolumn{1}{l|}{4287} & 4171 & \multicolumn{1}{l|}{$\ell=3$} \\ \hline \hline
MG-MLMC++, optimized target variances & \multicolumn{1}{l|}{181} & \multicolumn{1}{c|}{158} & \multicolumn{1}{l|}{143} & \multicolumn{1}{l|}{108} & 177 & \multicolumn{1}{l|}{$\ell=1$} \\ \cline{2-7} 
 & \multicolumn{1}{l|}{221} & \multicolumn{1}{l|}{221} & \multicolumn{1}{l|}{200} & \multicolumn{1}{l|}{148} & 111 & \multicolumn{1}{l|}{$\ell=2$} \\ \cline{2-7} 
 & \multicolumn{1}{l|}{278} & \multicolumn{1}{l|}{272} & \multicolumn{1}{l|}{243} & \multicolumn{1}{l|}{162} & 173 & \multicolumn{1}{l|}{$\ell =3$} \\ \hline
\end{tabular}
\caption{Number of stochastic samples for different masses at each level $\ell$ for deflated Hutchinson, MG-MLMC and MG-MLMC++, both with optimized target variances}
\label{tab:samples_nr}
\end{table}

As a supplementary information, Tab.~\ref{tab:samples_nr} reports the number of stochastic samples that were carried out on the different level differences. These numbers directly illustrate the variance reductions achieved in the different approaches. Each stochastic sample involves the solution of two linear systems (with matrices $D_\ell$ and $D_{\ell+1}$). These are done via multigrid and are thus quite efficient. This is why the numbers of stochastic samples do not reflect the total arithmetic cost of the methods, in which, in particular, performing the projections has a high cost when the deflating subspace becomes larger.  Interestingly, there is no visible dependence on the mass parameter for the MLMC approaches as was already observed in~\cite{frommer2022multilevel}. 

\section{Conclusion}

We have developed MG-MLMC++, a new trace estimator for the inverse which combines multigrid multilevel Monte Carlo with the recent Hutch++ approach. We have shown that the method outperforms other ones in trace computations for the Schwinger model. How to easily obtain a good choice for the number of vectors to use in the block power iteration is a subject of future research as is the application of our approach to the 4-dimensional (Wilson-) Dirac operator.

\bibliographystyle{JHEP}
\vspace*{-2mm}
\bibliography{lit}
\vspace*{-2mm}

\end{document}